\documentstyle[aps,epsfig,twocolumn]{revtex}

\begin{document}

\draft 

\title{$K^+$ versus $\Lambda$ flow in relativistic heavy-ion collisions}

\bigskip
\author{G.Q. Li and G.E. Brown}
\address{Department of Physics and Astronomy, State University of New
York at Stony Brook,\\
Stony Brook, New York 11794}
\maketitle
  
\begin{abstract}

We study $K^+$ and $\Lambda$ flow in heavy-ion collisions
at beam energies of about 2A GeV. We present our results
in both the `traditional' (i.e., in terms of the average transverse
momentum in the reaction plane) as well as `modern' 
(i.e., in terms of coefficients of the Fourier analysis
of azimuthal distributions) methods of flow analysis. We find 
significant differences between the $K^+$ and 
the $\Lambda$ flow: while the $\Lambda$ flow is basically
similar to that of nucleons, the $K^+$ flow almost
disappears. This difference is attributed chiefly to their
different mean field potentials in dense matter.
The comparisons with the experimental data, as well as
theoretical results from independent calculations, indicate
clearly the pivotal roles of both $K^+$ and $\Lambda$ medium
effects. We emphasize that similar experimental data from
independent collaborations are essential for the eventual
verification of these medium effects.

\end{abstract}

\pacs{25.75.Dw, 26.60.+c, 24.10.Lx}

\narrowtext

\section{Introduction}

Whether and how hadronic properties, such as their masses,
widths, and dispersion relations, are modified in hot and 
dense medium is a topic of great current interest. 
Of particular importance are the medium modifications of kaon 
properties, as they are related to both spontaneous and explicit
chiral symmetry breaking, and they are useful inputs for the
study of kaon condensation and neutron star properties
\cite{brown88a,thor94}. Since the pioneering work of Kaplan 
and Nelson \cite{kap86} on the possibility of kaon condensation 
in nuclear matter, a huge amount of theoretical effort has been 
devoted to the study of kaon properties in dense matter,
using such diversified approaches as chiral perturbation
theory \cite{brown87,wise91,brown94,lee95,kai95,waas96,lee96,kai97,waas97}, 
the Nambu$-$Jona-Lasinio model \cite{lutz94}, and SU(3) Walecka-type 
mean-field model \cite{sch94,knor95}. These calculations
indicate that $K^+$ feels a weak repulsive potential, arising
from the near cancellation of attractive scalar and repulsive
vector potentials. On the other hand, the $K^-$ feels a
strong attractive potential, since its vector potential
is attractive as well.

Closely related to this are the in-medium properties of hyperons,
which, like kaons, also carry explicitly the strangeness.
There also have been quite extensive theoretical
studies for hyperon (in particular $\Lambda$) properties
in nuclear matter. In Ref. \cite{speth94}, $\Lambda$
potential was calculated in the Dirac-Brueckner approach
using a boson-exchange model for the $\Lambda N$ interaction.
In Refs. \cite{rufa90,glen93} the SU(3) Walecka-type model
was used to study the hyperon properties. Generally, the
results from these theoretical studies are in agreement
with the empirical $\Lambda$ potential extracted
from the analysis of hypernucleus properties \cite{dov88,gib95},
namely, an attractive potential of about $-30$ MeV.

The analysis of $K^+$-nucleus scattering \cite{brown88} and hypernucleus
structure probes the properties of $K^+$ and $\Lambda$ 
at densities below normal nuclear matter densities $\rho_0$.
For the study of kaon condensation and the role of hyperons
in neutron stars \cite{gal97}, much higher densities are involved.
This can only be obtained by analysing heavy-ion collision 
data on various observables involving strangeness. 
Of particular interesting is a comparative and simultaneous
study of $K^+$ and $\Lambda$ observables, such as
their collective flow which is the focus of this paper.

In heavy-ion collisions at a few AGeV beam energies, 
the $K^+$ and $\Lambda$ are mostly produced together in
the so-called associated processes, such as
$BB\rightarrow B\Lambda K$ and $\pi B\rightarrow \Lambda K$,
where $B$ represents a baryon such as a nucleon or
a delta resonance. Without final-state interactions,
their momentum spectra are expected to be quite similar.
Thus any observed difference
in their flow patterns can mainly be attributed to the
difference in their final-state interactions. 
This includes both the rescattering with the environment (the
short-range correlation) and their propagation in
mean-field potentials (the long-range correlation).
In certain energy regions, the $K^+N$ and $\Lambda N$
cross sections are actually quite similar (see below),
thus the difference in their flow patterns may eventually
be traced back to the difference in their mean-field
potentials. 

The experimental and theoretical studies of collective flow
of various types have been an important component of
relativistic heavy-ion collisions \cite{qm,reis97}. Traditionally,
in-plane directed flow is characterized by the average transverse
momentum $\langle p_x \rangle$ as a function of the
rapidity $y$, based on the Danielewicz-Odyniec method \cite{dan85}.
This kind of `traditional' analysis has been carried
out mainly for heavy-ion collisions up to  Bevalac and SIS
energies (1-2A GeV) \cite{reis97}. Recently, there is a resurgence 
of the flow study in heavy-ion collisions at AGS (10A GeV)
\cite{e877a,e877b,e877c} and SPS (200A GeV) energies \cite{na49a,na49b}, 
using a method \cite{oll92,oll93,zhang96} based on the Fourier 
analysis of the azimuthal particle (energy) distributions. 
This `modern' method of flow analysis is less affected by the 
uncertainties in the reaction plane determination \cite{oll97}.

Kaon flow as a probe of the kaon potential in dense matter
was first propsed by Li {\it et al.} \cite{likoli95,liko95a}.
Lambda flow was studied by Li and Ko in Ref. \cite{liko96}.
Indeed, quite different flow patterns were predicted
for $K^+$ and $\Lambda$, because of their different mean-field
potentials. This has stimulated many experimental activities both
at SIS \cite{ritman,fopi,leif,hong} and AGS \cite{eos96,eos97,ogil98}.
So far, the experimental data from the FOPI collaboration
\cite{ritman,fopi,leif,hong} and the EOS collaboration \cite{eos96,eos97}
seem to support the scenario that the $K^+$ feels a 
weak repulsive potential, while the $\Lambda$ feels an attractive
potential, in nuclear matter.
 
This paper is an extension of Refs. \cite{likoli95,liko95a,liko96}.
We will first compare our results with existing experimental
data from the FOPI collaboration, and with
theoretical results from other independent calculations \cite{brat97,fae97}.
We will then report our predictions for Ru+Ru
collisions at 1.69A GeV, currently under investigation by
the FOPI collaboraton. Here we shall concentrate on the
centrality dependence of the $K^+$ and $\Lambda$ flow.
We will also present our predictions for Au+Au collisions
at 2A GeV, current under investigation by the E895
collaboration \cite{roy98,best97} (and possible by the E866 
collaboration \cite{ogil98}). Here we shall discuss
our results in terms of the Fourier coefficients of the
azimuthal distributions. 

This paper is arranged as follows. In Section II, we
review the relativistic transport model, the in-medium
properties of $K^+$ and $\Lambda$, and their production cross sections. 
The results for Ni+Ni and Ru+Ru are presented in Section III,
in terms of 'traditional' flow variables. The results for
Au+Au collisions are presented in Section IV, in terms
of `modern' flow variables. The paper ends with a short summary 
in Section IV.

\section{The relativistic transport model and strangeness production}

Heavy-ion collisions involve very complicated nonequilibrium
dynamics. One needs to use transport models in order to
extract from experimental data the information about 
in-medium properties of hadrons. In this work we will
use the relativistic transport model similar to that 
developed in Ref. \cite{ko87}. Instead of the usual linear 
or non-linear $\sigma$-$\omega$ models, we base our model on 
the effective chiral Lagrangian recently developed by Furstahl, 
Tang, and Serot \cite{fst}, which is derived using dimensional 
analysis, naturalness arguments, and provides a very good description 
of nuclear matter and finite nuclei. In the mean-field approximation, 
the energy density for the general case of asymmetric nuclear matter 
is given by
\begin{eqnarray}
\varepsilon _N& = & {2\over (2\pi )^3} \int _0^{K_{fp}} 
d{\bf k} \sqrt {{\bf k}^2+m_N^{*2}} +{2\over (2\pi )^3} 
\int _0^{K_{fn}} d{\bf k} \sqrt {{\bf k}^2+m_N^{*2}} \nonumber\\
 & + & W\rho +R {1\over 2}(\rho_p-\rho_n) -{1\over 2C_V^2}W^2
- {1\over 2C_\rho^2}R^2 + {1\over 2C_S^2}\Phi^2 \nonumber \\
 & +& {S^{\prime 2}\over 4C_S^2}d^2\left\{\left(1-{\Phi \over S^\prime}
\right)^{4/d}\left[{1\over d}{\rm ln}\left(1-{\Phi \over S^\prime}
\right) - {1\over 4}\right]+{1\over 4}\right\} \nonumber\\
 & -& {\xi\over 24}W^4 - {\eta \over 2C_V^2}{\Phi \over S^\prime}W^2. 
\end{eqnarray}
The nucleon effective mass $m_N^*$ is related to its scalar
field $\Phi$ by $m_N^*=m_N-\Phi$. $W$ and $R$ 
are the isospin-even and isospin-odd vector potentials,
respectively. The last three terms give the self-interactions of
the scalar field, the vector field, and the coupling between
them. The meaning and values of various parameters in Eq. (1)
can be found in \cite{fst}.

From the energy density of Eq. (1), we can also derive a 
relativistic transport model for heavy-ion collisions. 
At SIS energies, the colliding system consists mainly of 
nucleons, delta resonances, and pions. While medium effects 
on pions are neglected, nucleons and delta resonances propagate 
in a common mean-field potential according to the Hamilton 
equation of motion,
\begin{eqnarray}
{d{\bf x}\over dt} = {{\bf p}^*\over E^*}, \;\;\;
{d{\bf p}\over dt} = - \nabla _x (E^*+W),
\end{eqnarray}
where $E^*=\sqrt {{\bf p}^{*2} + m^{*2}}$.
These particles also undergo stochastic two-body
collisions, including both elastic and inelastic scattering. 

In heavy-ion collisions at incident energies considered in
this work, kaons and hyperons are produced together from pion-baryon and 
baryon-baryon collisions. For the former we use cross sections 
obtained in the resonance model by Tsushima {\it et al.} 
\cite{fae94}. For the latter the cross sections obtained in 
the one-boson-exchange model of Ref. \cite{liko95b,likoc98} 
are used. Both models describe the available experimental data 
very well. 

\begin{figure}[hbt]
\begin{center}
\centerline{\epsfig{file=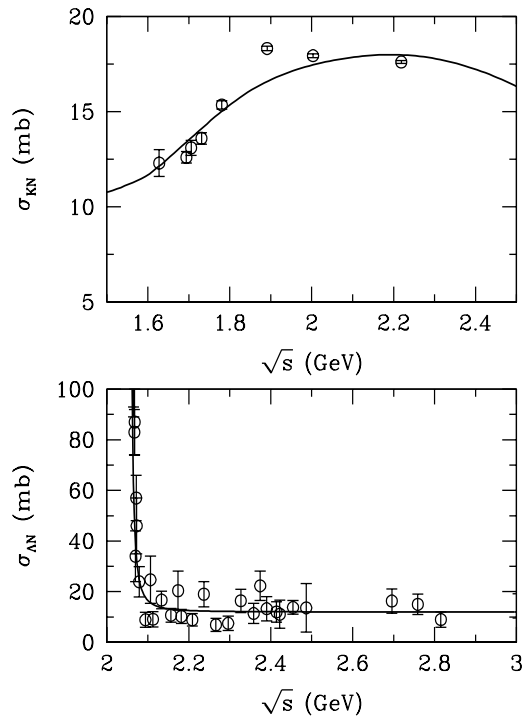,height=4.0in,width=3.7in}}
\caption{Kaon-nucleon and lambda-nucleon cross sections as
a function of the center-of-mass energy. \label{cross}}
\end{center}
\end{figure}

Particles produced in elementary hadron-hadron 
interactions in heavy-ion collisions cannot 
escape the environment freely. Instead, they
are subjected to strong final-state interactions.
For the kaon, because of strangeness conservation,
its scattering with nucleons at low energies is
dominated by elastic and pion production processes,
which do not affect its final yield but changes its momentum 
spectra. Similarly, at low-energies, the $\Lambda N$
collisions are also dominated by elastic scattering.
In Fig. \ref{cross}, we show the $K^+N$ and $\Lambda N$ cross sections.
The solid lines are our parameterizations used in the
transport model, while the circles are experimental
data from Ref. \cite{xdata}. It is seen that except at very low
beam momenta, the $K^+N$ and $\Lambda N$ cross
sections are quite similar (of the order of 10-15 mb).

We will consider two scenarios for kaon properties in the nuclear
medium, one with and one without medium modification.  
From the chiral Lagrangian the kaon in-medium energies
can be written as \cite{lilee97b}
\begin{eqnarray}
\omega _K=\left[m_K^2+{\bf k}^2-a_K\rho_S
+(b_K \rho )^2\right]^{1/2} + b_K \rho 
\end{eqnarray}
where $b_K=3/(8f_\pi^2)\approx 0.333$ GeVfm$^3$, 
$a_K$ is the parameter that determines the strength of the 
attractive scalar potential for the kaon. If one considers only the
Kaplan-Nelson term, then $a_K=\Sigma _{KN}/f_\pi ^2$.
In the same order, there is also the range term which 
reduces the scalar attraction.
Since the exact value of $\Sigma _{KN}$ and the size of
the higher-order corrections are still under intensive
debate, we take the point of view that $a_{K}$ can be treated 
as free parameter and try to constrain it from the experimental 
observables in heavy-ion collisions. In Ref. \cite{lilee97a,lilee97b} 
it was found that $a_K\approx 0.22$ GeV$^2$fm$^3$ provide a good 
description of $K^+$ spectra in Ni+Ni collisions. This value will
be used in this work as well. Actually, since $KN$ interaction
is relatively weak, the kaon potential in nuclear matter, at least
at low densities, can also be obtained from the $KN$ scattering
length in free space using the impulse approxation. 
In Fig. 2 we show the kaon potential as a function of density,
which is defined as
\begin{eqnarray} 
U_K({\bf k}, \rho ) = \omega _K - (m_K^2+{\bf k}^2)^{1/2}.
\end{eqnarray}
The open circle in the figure is the kaon potential at $\rho_0$ 
obtained in the impulse approximation.

It is well-known that the quark counting
rule applies approximately to the $\Lambda$ potential. Thus
the $\Lambda$ potential is about 2/3 of that of nucleon:
\begin{eqnarray}
\Phi _\Lambda = 2/3 \Phi , ~~ W_\Lambda = 2/3 W.
\end{eqnarray}
The $\Lambda$ optical model potential can then be defined
as
\begin{eqnarray}
U_\Lambda ({\bf k}, \rho ) 
=\left((m_\Lambda -\Phi _\Lambda)^2+{\bf k}^2\right)^{1/2}
+W_\Lambda - \left(m_\Lambda ^2+{\bf k}^2\right)^{1/2}.
\end{eqnarray}
The $\Lambda$ potential obtained in this way is also shown in
Fig. 2. The solid circle represents its potential
$\rho _0$ extracted from the structure of hypernuclei.
It is seen that for the nuclear density region relevant
for this study, the $\Lambda$ potential is attractive,
while that of the $K^+$ is repulsive.
This difference will affect their final momentum 
distributions, through their propagation in
these potentials. 
The Hamilton equations of motion for kaon and lambda are very
similar to those for nucleons \cite{liko95a,liko96},
\begin{eqnarray}
{d{\bf r}\over dt} = {{\bf k}\over \omega _{K} - b_k\rho _N},
\;\; {d{\bf k}\over dt} = - \nabla _x U_{K},
\end{eqnarray}
\begin{eqnarray}
{d{\bf r}\over dt} = {{\bf k}\over E_{\Lambda}^*},
\;\; {d{\bf k}\over dt} = - \nabla _x U_\Lambda,
\end{eqnarray}
where $E^*_\Lambda =\left((m_\Lambda -\Phi _\Lambda)^2+{\bf k}^2\right)^{1/2}$. 

\begin{figure}[hbt]
\begin{center}
\centerline{\epsfig{file=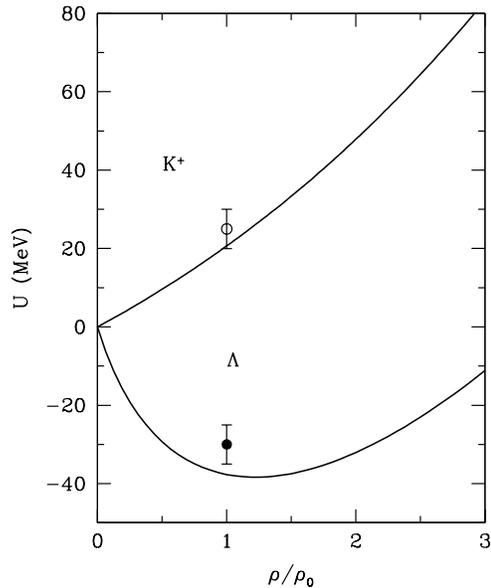,height=3.5in,width=3.5in}}
\caption{The kaon and lambda potentials used in this work. The
circles are their `empirical' potentials at normal nuclear
matter density. \label{pot}}
\end{center}
\end{figure}

\section{$K^+$ versus $\Lambda$ flow in `traditional' analysis}

\begin{figure}[hbt]
\begin{center}
\centerline{\epsfig{file=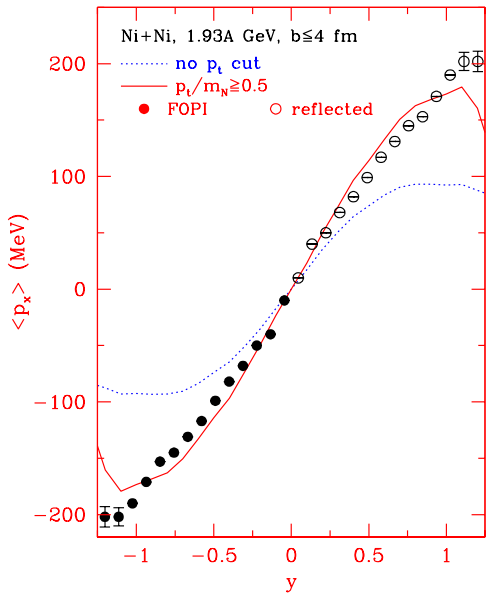,height=3.5in,width=3.5in}}
\caption{Comparison of proton flow with experimental data
from the FOPI collaboration \protect\cite{ritman}. The role
of the transverse momentum cut is highlighted. \label{protflow}}
\end{center}
\end{figure}

In this section we discuss the $K^+$ and $\Lambda$ flow in Ni+Ni
and Ru+Ru collisions, in terms of `traditional' flow analysis, namely,
the average transverse momentum and the associated flow parameters.
We will compare our results with available data from the FOPI and
EOS collaborations. Before this, we first compare our results
for proton flow with the FOPI data \cite{ritman,fopi,leif,hong}.
This is done in Fig. \ref{protflow}, where the solid and dotted curves
are our results with and without the transverse momentum cut,
which was applied in the experimental data. The transverse 
momentum cut enhances significantly the flow strength, and our
results include this cut are in good agreement with the data.
This is important, as all the following discussions about
$K^+$ and $\Lambda$ flow are presented with respect to the
nucleon flow.

\subsection{Ni+Ni collisions}

\begin{figure}[hbt]
\begin{center}
\centerline{\epsfig{file=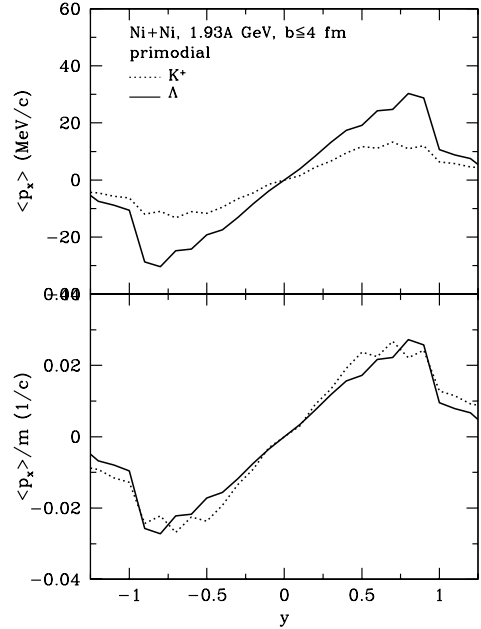,height=3.5in,width=3.5in}}
\caption{$K^+$ and $\Lambda$ flow in central Ni+Ni collisions
at 1.93A GeV. The upper window shows the average transverse
momentum, while the lower window shows the velocity (normalized by
their respective masses). The results shown are for the case
without any final-state interactions. \label{klfp}} 
\end{center}
\end{figure}

As mentioned in the Introduction, since the $K^+$ and $\Lambda$ are
produced together in associated processes, their flow pattern
should be very similar, if there were no any final-state interactions
for these particles. The flow of primodial kaons and lambda hyperons
should reflect the collective flow of baryon-baryon and
pion-baryon pairs from which they are produced. The
results are shown in Fig. \ref{klfp}. As can be seen, both $K^+$
and $\Lambda$ show positive flow, in the same direction as
that of nucleons. In terms of the average transverse momentum,
the $\Lambda$ flow is stronger than that of $K^+$, because it
is heavier. Actually, their transverse velocities are quite similar,
as shown in the lower window of the figure. This velocity is acquired
from the Lorentz boost in the direct of nucleon flow. Both the
$K^+$ and $\Lambda$ flow, without any final-state interactions, are
much weaker than that of nucleon (see the dotted line in Fig. \ref{protflow}.
This statement can be quantified by introducing the so-called
flow parameter $F$, defined as the slope parameter of the transverse
momentum curve at the mid-rapidity,
\begin{eqnarray}
F = {d\langle p_x \rangle \over dy} | _{y=0}.
\end{eqnarray}
While the nucleon flow parameter is about 150 MeV/c, the flow parameters of
the primodial $K^+$ and $\Lambda$ are 20 and 40 MeV/c, respectively.

\begin{figure}[hbt]
\begin{center}
\centerline{\epsfig{file=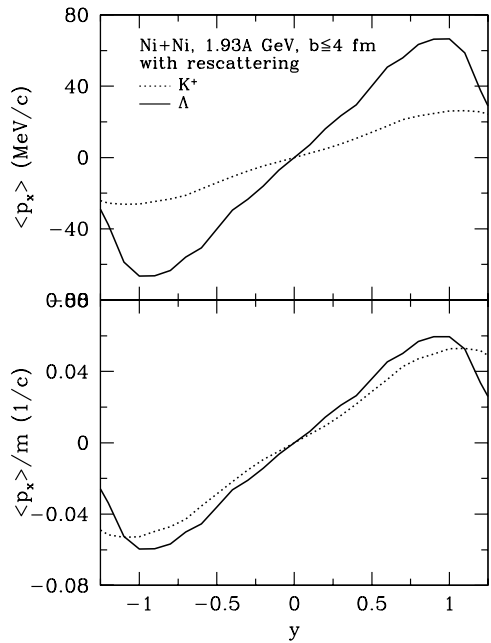,height=3.5in,width=3.5in}}
\caption{Same as Fig. \protect\ref{klfp}, including 
rescattering. \label{klfs}} 
\end{center}
\end{figure}

Including the $KN$ and $\Lambda N$ rescattering increases their
flow in the direction of nucleons, as shown in Fig. \ref{klfs}. 
This enhancement is mainly due to the
thermalization effects, which increase the average
momenta of kaons and lambda hyperons. At the beam energies considered
here, which are only slightly above the production threshold,
the produced kaons and lambda hyperons usually have small
momenta. By rescattering with energetic nucleons, their
momenta increase, as does their flow strength. The $K^+$ flow
parameter increases from 20 to 40 MeV, while the $\Lambda$
flow parameter increases from 40 to 75 MeV.
Their flow velocities are still about the same, since
their rescattering cross sections are not very different.

\begin{figure}[hbt]
\begin{center}
\centerline{\epsfig{file=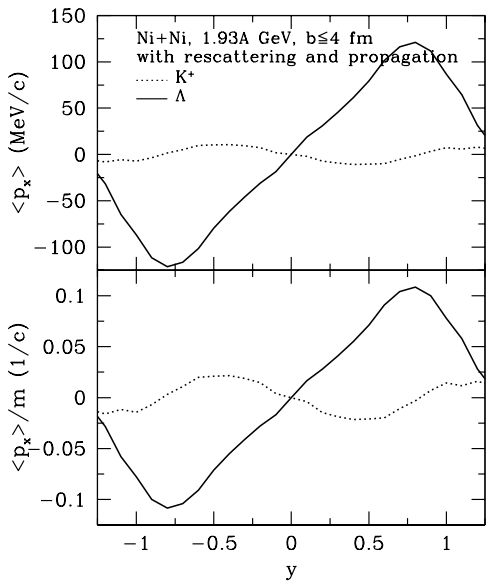,height=3.5in,width=3.5in}}
\caption{Same as Fig. \protect\ref{klfp}, including 
rescattering and propagation in mean field potentials. \label{klfm}} 
\end{center}
\end{figure}

\begin{figure}[hbt]
\begin{center}
\centerline{\epsfig{file=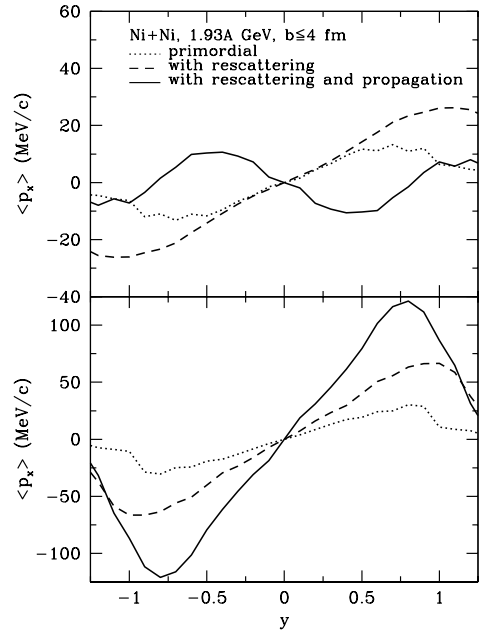,height=3.5in,width=3.5in}}
\caption{$K^+$ and $\Lambda$ flow in central Ni+Ni collisions
at 1.93A GeV. The effects of final-state interactions are
hightlighted. \label{klftot}} 
\end{center}
\end{figure}

Finally the results for $K^+$ and $\Lambda$ flow including both
rescattering and propagation in mean-field potentials are
shown in Fig. \ref{klfm}. Since the $K^+$ potential is 
repulsive, kaons are pushed away from nucleons, leading
to the near disappearance of the flow signal. The flow 
parameter turns from positive into negative, and is about -15 MeV. On the 
other hand, the $\Lambda$ potential is attractive, and lambda hyperons 
are pulled towards nucleons, leading to the further 
enhancement of its flow in the direction of nucleon flow.
The $\Lambda$ flow parameter increases to about 130 MeV, close to
that of nucleons. Therefore, the difference in their mean-field
potentials lead to substantially different flow pattern for
$K^+$ and $\Lambda$, although they are produced together
in the same processes, and their rescattering cross sections
are not very difference. To see more clearly the
effects of final-state interactions, these results
are summarized in Fig. \ref{klftot}, where the upper
window shows those for $K^+$ and the lower window for
$\Lambda$.

\begin{figure}[hbt]
\begin{center}
\centerline{\epsfig{file=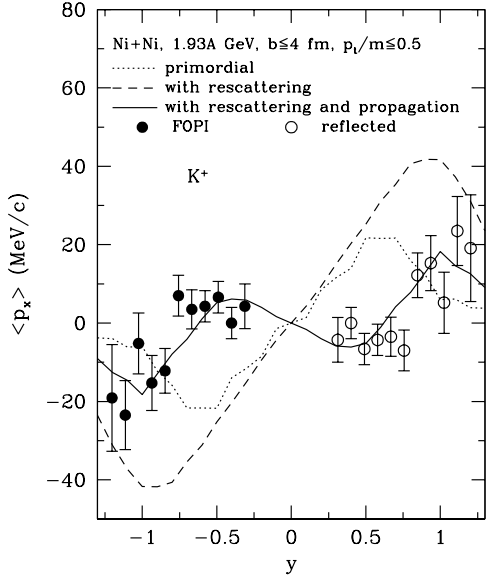,height=3.5in,width=3.5in}}
\caption{Comparison of $K^+$ flow with experimental data
from the FOPI collaboration \protect\cite{ritman}. 
\label{kflow}}
\end{center}
\end{figure}

Since the proposal of kaon flow as a signal of kaon potential
in Ref. \cite{likoli95}, the FOPI collaboration measured
both $K^+$ and $\Lambda$ flow in Ni+Ni collisions at 1.93A GeV
\cite{ritman,fopi,leif,hong}, while the EOS collaboration
measured $\Lambda$ flow in Ni+Cu collisions at 2A GeV \cite{eos96,eos97}.
The comparison of our results with the FOPI data for $K^+$ flow are
shown in Fig. \ref{kflow}. As in the experimental data, our
results include a transverse momentum cut of $p_t/m_K > 0.5$,
which increases the flow signal in the direction of the nucleon
flow. Clearly, without the kaon medium effects, namely, the 
propagation in the repulsive mean-field
potential, the $K^+$ flow is positive, and is in complete disagreement
with the experimental data, which show a small antiflow in the 
mid-rapidity region. Including the effects of the repulsive 
potential, the kaons are pushed away from the nucleon, and our
results are now in very good agreement with data. This indicates
that the kaon potential as determined by Eq. (3) is reasonable.

\begin{figure}[hbt]
\begin{center}
\centerline{\epsfig{file=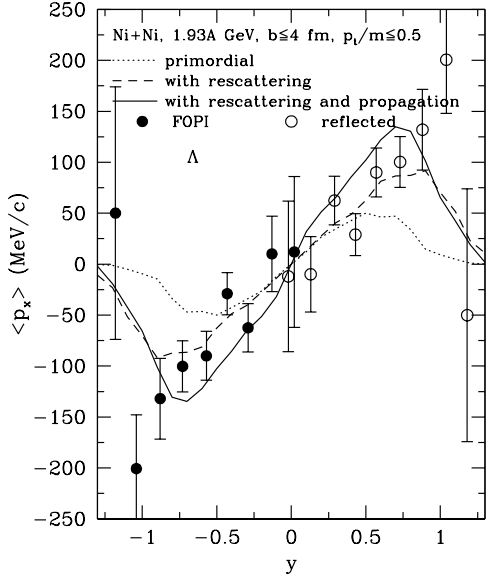,height=3.5in,width=3.5in}}
\caption{Comparison of $\Lambda$ flow with experimental data
from the FOPI collaboration \protect\cite{ritman}. 
\label{lflow}}
\end{center}
\end{figure}

The comparsion of our results for $\Lambda$ flow in the same system
are shown in Fig. \ref{lflow}. Experimental data \cite{ritman}, 
though with quite large statistical uncertainties, clearly show that
the $\Lambda$ flow is in the same direction as nucleons, with
a strength quite similar to that of nucleons. A similar observation
has also been made by the EOS collaboration \cite{eos96,eos97}.
Such a strong positive flow can be achieved only
after the inclusion of the $\Lambda$ final-state interaction.
An accurate determination of the role of the mean-field potentials, however,
requires much improved data, especially near the mid-rapidity.
 
\begin{figure}[hbt]
\begin{center}
\centerline{\epsfig{file=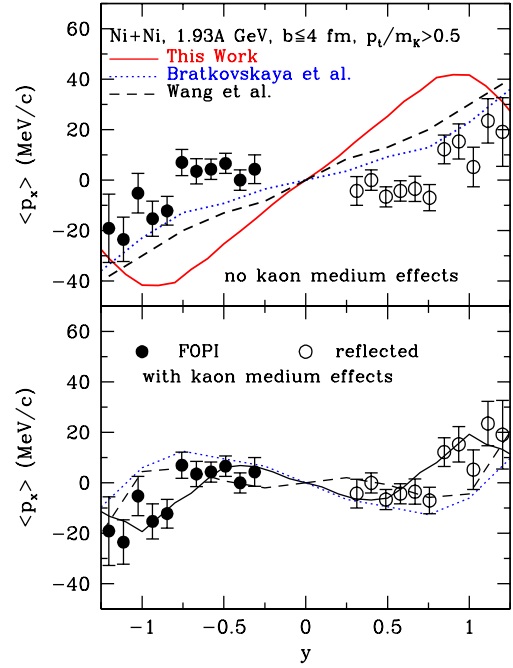,height=4.0in,width=3.9in}}
\caption{Comparison of $K^+$ flow from three independent 
calculations with experimental data from the FOPI collaboration 
\protect\cite{ritman}. 
\label{kflowcom}}
\end{center}
\end{figure}

Theoretically, kaon flow in Ni+Ni collisions has also been studied 
by several independent groups, using different dynamical models
such as the Hadron-String Dynamics \cite{brat97}, the Quantum-Molecular
Dynamics (QMD) \cite{fae97}, and the Relativistic Boltzmann-Uehling-Uhlenbeck
(RBUU) \cite{li98}. The results from these calculations are
qualitatively similar to ours, as summarized in Fig. \ref{kflowcom}.
Without kaon medium effects, all the calculations predict
a positve flow signal for kaons, and thus in disagreement with the
experimental data (see upper window of the figure).
Including the repulsive kaon potential, the kaon flow turns
into a small antiflow, and all the three results are in good 
agreement with the data, as shown in the lower window of the
figure.

\subsection{Ru+Ru collisions: predictions}

\begin{figure}[hbt]
\begin{center}
\centerline{\epsfig{file=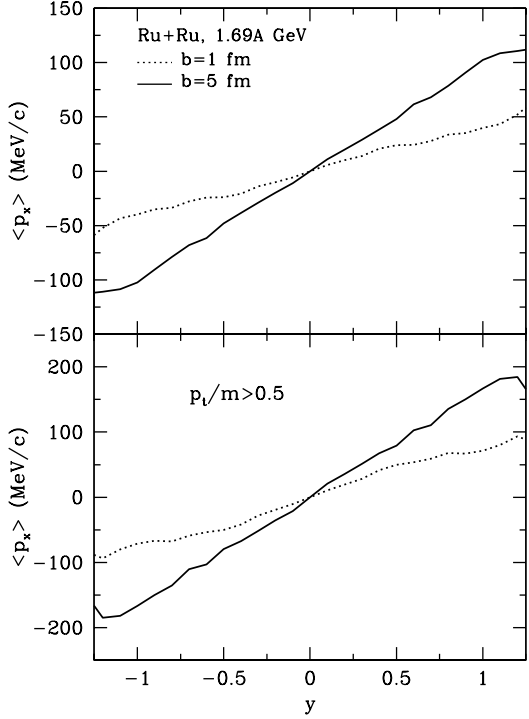,height=4.0in,width=3.9in}}
\caption{Proton flow in Ru+Ru collisions at two impact parameters,
with and without the transverse momentum cut. 
\label{pfru}}
\end{center}
\end{figure}

We expect that the $K^+$ and $\Lambda$ flow in Ru+Ru collisions
at 1.69A GeV to be very similar to that in Ni+Ni collisions
at 1.93A GeV. To study the centrality dependence of the these
flow patterns, we choose two impact parameters b=1 anf 5 fm,
corresponding to central and semi-central collisions. As a
reference, we show in Fig. \ref{pfru} the proton flow with
and without the tranverse momentum cut of $p_t/m >0.5$. This
cut is included to facilitate comparisons with future 
FOPI data. As expected, the proton flow is larger in semi-central
collisions than in central ones.

\begin{figure}[hbt]
\begin{center}
\centerline{\epsfig{file=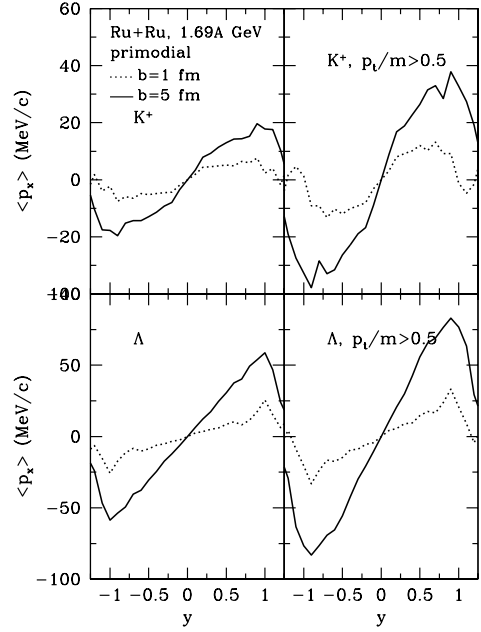,height=3.5in,width=3.5in}}
\caption{$K^+$ and $\Lambda$ flow in Ru+Ru collisions
at 1.69A GeV and two impact parameters. The upper windows show 
the results for $K^+$, and the lower windows for $\Lambda$.
The results in right windows include a transverse momentum
cut of $p_t/m >0.5$. \label{klfpru}}
\end{center}
\end{figure}

\begin{figure}[hbt]
\begin{center}
\centerline{\epsfig{file=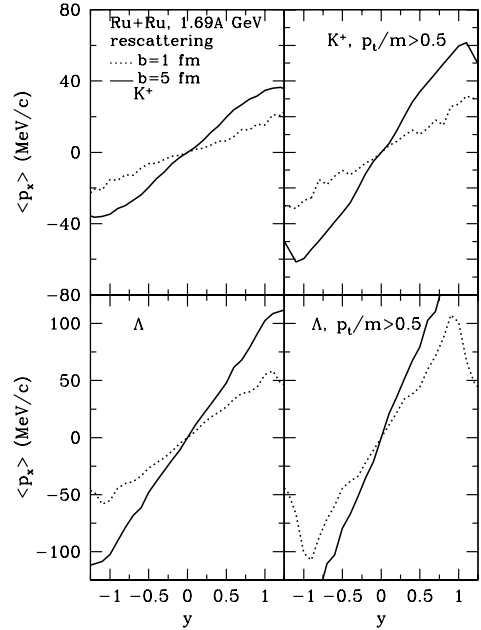,height=3.5in,width=3.5in}}
\caption{Same as Fig. \protect\ref{klfpru}, including 
rescattering. \label{klfsru}} 
\end{center}
\end{figure}

The results for the primodial $K^+$ and $\Lambda$ flow are shown 
in Fig. \ref{klfpru}. The results in the right panel include
a transverse momentum cut of $p_t/m>0.5$. In this case both the
kaons and lambda hyperons flow in the same direction as
the nucleons, but with a substantially smaller flow velocity.
Just like nucleons, the flow of primodial kaons and hyperons
is stronger in semicentral collisions than in central collisions.
The results including the $KN$ and $\Lambda N$ rescattering are
shown in Fig. \ref{klfsru}. Again, the rescattering increases
both the $K^+$ and $\Lambda$ flow in the direction of nucleon
flow, and the rescattering effects are about the same for the
central and semicentral collisions.

\begin{figure}[hbt]
\begin{center}
\centerline{\epsfig{file=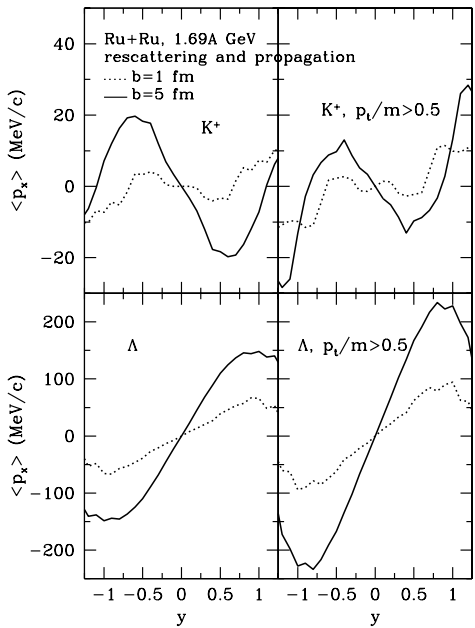,height=3.5in,width=3.5in}}
\caption{Same as Fig. \protect\ref{klfpru}, including 
rescattering and propagation in mean-field potentials. 
\label{klfmru}} 
\end{center}
\end{figure}

Finally, the results including both the rescattering and
propagation in the mean-field potentials are shown in
Fig. \ref{klfmru}. Kaons are pushed away from nucleons by
their repulisive potential, and lambda hyperons are pulled
towards nucleon by their attractive potential, enhancing 
further its flow strength. The mean-field effects are seen to
be stronger in semi-central collisions than in central
ones. Although compression might be larger in central
collisions, the density gradients, which deterimine the
forces entering Eqs. (7) and (8), is larger in semi-central
collisions, because of the existence of the spectator matter.
This might explain why kaons in the semi-central collisions
are pushed even further away from nucleons than those in
central collisions, although the kaons in the former case
have a larger positive flow than in the latter case (see
Fig. \ref{klfsru}) before the propagation in the mean-field
potential sets in. After including the transverse 
momentum cut, the kaons are clearly anticorrelated with
nucleons in semicentral collisions, with a flow parameter
of about -20 MeV. On the other hand, at least near
mid-rapidity, the kaons in the central collisions are
seen to be correlated with nucleons, with a flow parameter
of about 10 MeV. The study of the centrality dependence
of the kaon flow by the FOPI collaboration \cite{herr} will
shed further light on the question of kaon medium 
effects in heavy-ion collisions.
 
\section{$K^+$ versus $\Lambda$ flow in `modern' analysis}

Recently, the study of collective flow has been quite active at
higher AGS \cite{e877a,e877b,e877c} and SPS \cite{na49a,na49b}
energies. In these studies, the directed flow and the elliptic
flow are usually characterized by the first (dipole) and second
(quadrupole) moments of the Fourier analysis of the
azimuthal distribution. Most generally and completely,
an event in a relativistic heavy-ion collisions is 
characterized by the triple-differential cross sections
of all the particles,
\begin{eqnarray}
E{d^3N\over d^3p}& = &{d^3N\over p_tdp_t dy d\phi} \nonumber\\
 & = & {d^2N\over p_tdp_t dy} {1\over 2\pi} \left( 1
+ \sum _{n=1} 2v_n {\rm cos} (n\phi ) \right),
\end{eqnarray}
where an expansion in terms of the azimuthal angle $\phi$ is 
introduced. In practice, this expansion can be truncated at $n=2$, so that
\begin{eqnarray}
E{d^3N\over d^3p}\approx {d^2N\over p_tdp_t dy} {1\over 2\pi} \left( 1
+  2v_1 {\rm cos} \phi + 2v_2 {\rm cos}(2\phi ) \right).
\end{eqnarray}
Here $v_1$ reflects the relative abundance of particles in the
positive $x$-axis versus that in the negative direction (directed
flow), whereas $v_2$ measures the relative abundance of
particles in the reaction plane versus that of out-of the reaction
plane. Generally, both $v_1$ and $v_2$ depend on the transverse
momentum $p_t$ and rapidity $y$. Integrating over transverse
momentum, they can then be presented as a function of the
rapidity, in the same way as the average transverse momentum
is plotted as a function of the rapidity in the `traditional'
analysis. In this section we shall concentrate on the 
the directed flow of $K^+$ and $\Lambda$, as measured by 
the first Fourier moment $v_1$,
in Au+Au collisions at 2A GeV.

\begin{figure}[hbt]
\begin{center}
\centerline{\epsfig{file=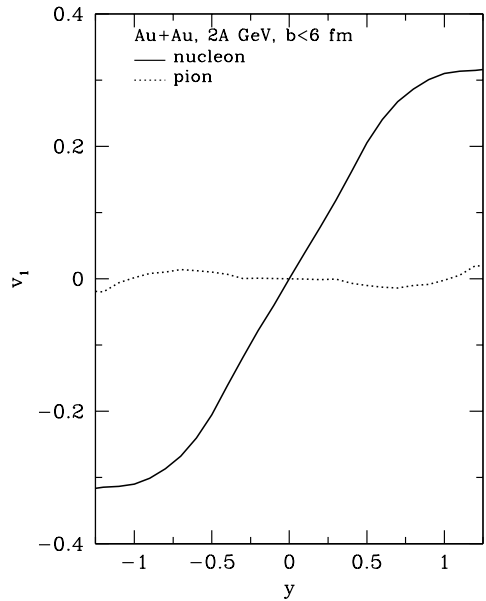,height=3.5in,width=3.5in}}
\caption{The rapidity dependence of the first moment ($v_1$) of
proton and pions in Au+Au collisions at 2A GeV and $b< 6$ fm. 
\label{protv1}} 
\end{center}
\end{figure}

We first show in Fig. \ref{protv1} the proton and pion flow for
impact parameter $b<6$ fm, which corresponds approximately
to the centrality selection of the E895 collaboration \cite{roy98}. 
We see a very strong positive flow for protons, which reaches
about 0.3 nearly the projectile (target) rapidity. On the
other hand, the pions show a very weak antiflow with respect
to nucleons. This anticorrelation arises from the strong
absorption of pions by spectator nucleons.
 
\begin{figure}[hbt]
\begin{center}
\centerline{\epsfig{file=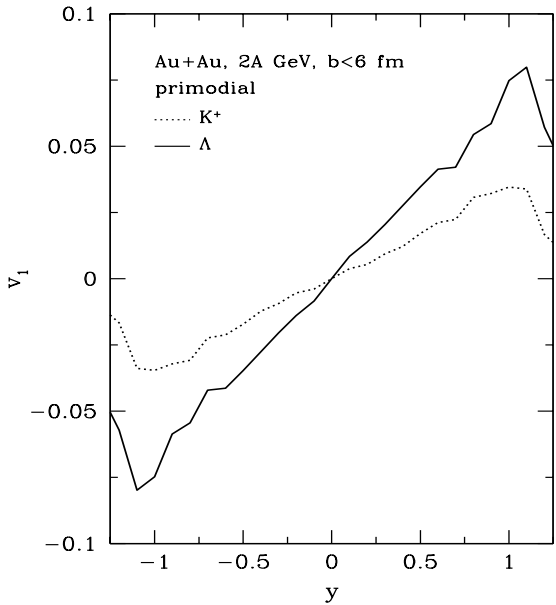,height=3.5in,width=4in}}
\caption{The rapidity dependence of the first moment ($v_1$) of
$K^+$ and $\Lambda$ in Au+Au collisions at 2A GeV. 
The results shown are for the case without any final-state
interactions.
\label{klv1p}} 
\end{center}
\end{figure}

\begin{figure}[hbt]
\begin{center}
\centerline{\epsfig{file=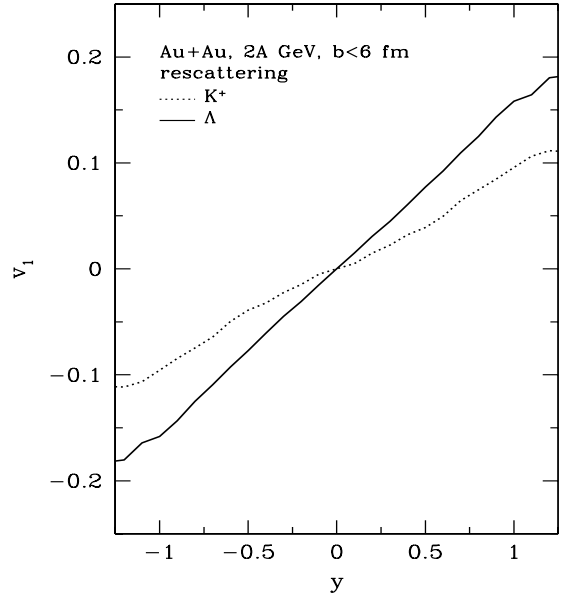,height=3.5in,width=4in}}
\caption{Same as Fig. \protect\ref{klv1p}, with rescattering.
\label{klv1s}} 
\end{center}
\end{figure}

\begin{figure}[hbt]
\begin{center}
\centerline{\epsfig{file=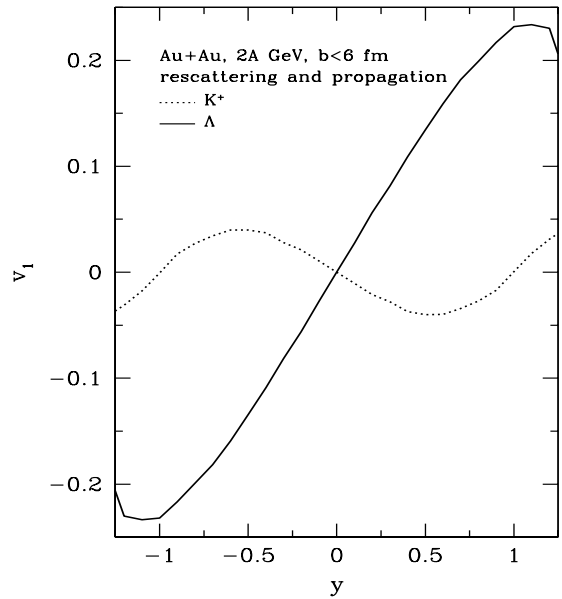,height=3.5in,width=4in}}
\caption{Same as Fig. \protect\ref{klv1p}, with rescattering
and propagation in the mean-field potentials.
\label{klv1m}} 
\end{center}
\end{figure}

The results for $K^+$ and $\Lambda$ flow in these collisions
are shown in Figs. \ref{klv1p}, \ref{klv1s}, and \ref{klv1m},
the the scenarios without final-state interaction, with 
rescattering, and with rescattering and propagation in mean-field 
potential, respectively. No transverse momentum acceptance has been
applied. 

Without any final-state interactions, both $K^+$ and $\Lambda$
show positive flow, in the same direction as nucleon, but
with a much smaller $v_1$. Also, if we normalize these moments
by their respective masses, we find that the normalized first
moments of $K^+$ and $\Lambda$ are about the same, reflecting the fact
that their flow velocities are similar, since they are produced
from the same sources. The rescattering with nucleons is seen
to increase the first moments of $K^+$ and $\Lambda$ in the direction
of nucleon flow, just as it enhances their average transverse
momenta. The flow patterns of the kaons and lambda hyperons
are still quite similar after the includsion of their rescattering
with nucleons. 

Finally, after the inclusion of the propagations in their
mean-field potentials, the flow patterns of $K^+$ and
$\Lambda$ become significantly different.
Kaons are pushed away from nucleons, leading to an anticorrelation
with nucleons. From the $v_1-y$ plot, we can also define
a so-called flow parameter, $F_{v_1}$, as the slope of this plot
at mid-rapidity. Clearly, the nucleons have a positive flow 
parameter, while the kaons show a negative flow parameter,
after the inclusion of the mean-field effects. On the
other hand, the lambda hyperons are pulled towards nucleons,
leading to an enhancement of its flow in the same direction
as nucleons. Its flow parameter after including these final-state
interaction is quite similar to that of nucleons.
Basically the $v_1$ versus $y$ plot provides quite similar information
as the $\langle p_x \rangle $ versus $y$ plot. The advantage
of working with $v_1$ is that one can obtain further information
by analysing its tranverse momentum (mass) dependence \cite{e877c}.
  
\section{Summary}

We studied $K^+$ and $\Lambda$ flow in heavy-ion collisions
at beam energies of about 2A GeV, using the relativistic
transport model that includes the strangeness degrees of
freedow explicitly. We find that, without any final-state
interactions, both the $K^+$ and $\Lambda$ flow in the 
same direction as nucleons, but with much smaller flow
velocities, which are quite similar for kaons and lambda hyperons.
This small flow velocity reflects the collective flow of
the baryon-baryon and pion-baryon pairs from which kaons
and lambda hyperons are produced.  The inclusion of their
rescattering with nucleons enhances the flow of $K^+$ and $\Lambda$
in the direction of nucleons, as a results of thermalization
effects. The flow velocities of kaons and lambda hyperons
in this case are still quite similar, and smaller than that
of nucleons. We find, furthermore, that the propagation in
their mean-field potentials leads to quite different
flow patterns for $K^+$ and $\Lambda$. Kaons are pushed away
from nucleons by their repulsive potential, and lambda 
hyperons are pulled towards nucleons by their attractive
potentials. This leads to the small antiflow of kaons with 
respect to nucleons, and the flow of lambda hyperons that
is very close to that of nucleons. The comparison of our results
with experimental data from the FOPI collaboration for Ni+Ni
collisions, and with independent theoretical calculations,
indicate clearly the important role of mean-field potentials.
We also presented predictions for Ru+Ru collisions 
in terms of average transverse momentum $\langle p_x \rangle$, 
and for Au+Au collisions in terms of the first moment
$v_1$ of the Fourier analysis of azimuthal particle distributions.

\vskip 0.5cm

We are grateful to N. Herrmann and R. Lacey
useful discussions. This work is supported in part by the 
Department of Energy under Grant No. DE-FG02-88ER40388.

\end{document}